\documentclass[11pt,a4paper]{article}

\def\AmSTeX{\leavevmode\hbox{$\mathcal A\kern-.2em\lower.376ex%
        \hbox{$\mathcal M$}\kern-.2em\mathcal S$-\TeX}}
\usepackage{graphicx,subfigure,boxedminipage}
\usepackage{srcltx}
\usepackage{ifpdf}
\usepackage{upgreek} 
\usepackage{amsmath}        
\usepackage{amssymb}
\usepackage{mathrsfs}       
\usepackage{bm}             
\usepackage[amsmath,thmmarks,hyperref]{ntheorem}
\usepackage{slashed}
\usepackage{color}
\newif\ifpdf \pdftrue
\ifx\pdfoutput\undefined \pdffalse \fi \ifx\pdfoutput\relax
\pdffalse \fi
\ifx\texonly\undefined\let\texonly\relax\fi
\ifx\endtexonly\undefined\let\endtexonly\relax\fi \texonly
  \let\htmlonly\iffalse
  \let\endhtmlonly\fi
\endtexonly

\htmlonly
  
\endhtmlonly
\texonly
  \usepackage{makeidx}

  \usepackage{multicol}
  

\endtexonly
\title{}
\author{\thanks{}}
\date{}
\begin{document}
\title{The Quark Propagator in a Truncation Scheme beyond the Rainbow Approximation}
\author{Hui-Feng Fu$^{1,2}$~\footnote{huifengfu@tsinghua.edu.cn}~~and Qing Wang$^{1,2,3}$~\footnote{wangq@mail.tsinghua.edu.cn}\\
{\it \small 1. Department of Physics, Tsinghua University, Beijing 100084, People’s Republic of China}\\
{\it \small 2. Center for High Energy Physics, Tsinghua University, Beijing 100084, People’s Republic of China}\\
{\it \small 3. Collaborative Innovation Center of Quantum Matter, Beijing 100084, People’s Republic of China}
 }

\maketitle

\baselineskip=20pt
\begin{abstract}
The quark propagator is studied under a truncation scheme beyond the rainbow approximation by dressing the quark-gluon vertex non-perturbatively. It is found that, in the chiral limit with dynamical symmetry breaking, the dynamical quark mass and the quark condensate are significantly enhanced due to the non-Abelian contribution arising from the three-gluon interaction compared to those under the rainbow approximation; and the critical strength of the dynamical chiral symmetry breaking is much lowered. The Abelian contribution is much smaller than the non-Abelian contribution. A technical issue on removing the ultraviolet divergences including the overlapping divergences is discussed.
\end{abstract}

\section{Introduction}
Understanding non-perturbative phenomena of QCD, such as the dynamical chiral symmetry breaking (DCSB), confinement and the low-lying hadrons spectra, is one of the most challenging tasks in theoretical physics. Continuous efforts have been made in this area. The quark propagator, a basic ingredient of QCD, has close connections to these non-perturbative phenomena (the dynamical generation of the running quark mass from a chiral symmetric Lagrangian indicates DCSB; confinement is related to the analytic property of the quark propagator; the quark propagator plays an important part through the wave functions for hadrons), thus deserves a non-perturbative investigation. The Dyson-Schwinger equations (DSEs) are fundamental coupled integral equations for the Green functions of the underlying quantum field theory, thus provide a natural way for such a study.

The DSE for the quark propagator is usually called the gap equation. It involves the gluon propagator and the quark-gluon vertex (QGV) as parts of the integrand. A typical approximation to truncate the DSEs is the rainbow approximation (RA), which replaces the quark-gluon vertex (QGV) with the bare one $\gamma_\mu$ (with a color matrix), and has been extensively employed~\cite{AJ,MJ,RW,MR,MT,CM,MRRS,Kra}. (In some of these works, the Bethe-Salpeter equation (BSE) under the ladder approximation was taken to study meson properties, and the rainbow-ladder approximation forms a chiral symmetry preserving truncation scheme). In spite of the achievements have been made with this truncation scheme, the need of going beyond the rainbow (BR) approximation has already been recognized~\cite{BDTR,FW,WF,RHK}. Indeed, since the full quark-gluon vertex includes 12 independent form factors embodying the dynamical information of the underlying theory, replacing the full vertex with $\gamma_\mu$ implies loss of this dynamical information. Especially, the gluon self-interaction contributions (to the quark-gluon vertex), which characterize the non-Abelian feature of QCD, will be lost if one takes the rainbow approximation (the vertex would be the same up to a color matrix for QED and QCD under the RA).

Going beyond the rainbow approximation can be done either by using the gauge invariance (and other constraints) to constrain the fermion-gauge-boson vertex and modeling it, or by dressing the vertex according to its own equation. The former method was first developed in QED by applying the Ward-Takahashi identity (WTI) to construct the fermion-gauge-boson vertex in terms of the fermion propagator. The BC vertex~\cite{BC} and the CP vertex~\cite{CP1} formed in this manner are extensively applied~\cite{CP2,HW2,HW,HMR,Cha,CR}. For QCD, the Slavnov-Taylor identity (STI) for the QGV, which also involves the ghost propagator and the ghost-quark scattering kernel, is used to model the quark-gluon vertex~\cite{FA,AP,Roj}. These identities only constrain the longitudinal part of the fermion-gauge-boson vertex, while the transverse part cannot be totally fixed. So along this direction, further efforts have been made~\cite{He,QCLRS}. The vertices modeled in this manner, reflect the symmetries and/or other physical requirements, however, how the dynamics of the underlying theory dresses the vertex is hard to be traced. For example, in a non-Abelian theory, how the gauge boson's self interaction affects the dressed fermion-gauge-boson vertex is hidden. Another method, dressing the vertex by using its own equation, offers an opportunity to discuss this kind of issues. In QCD, the QGV satisfies its own DSEs and the equation of motion \cite{AFLS}. In a certain sense of loop expansion, where the propagators are all fully dressed (the situation about the vertices in the expansion are complicated and will be specified in the next section), the three-gluon interaction comes into the dressed QGV at one loop level as the ``next order correction" to the bare vertex, thus provide a good opportunity to directly test the effects of this non-Abelian type interaction. Another correction comes from a loop diagram including only the quark-gluon vertices. Because this type of diagram also appears in the Abelian theory--QED, the corresponding diagram is referred as the Abelian diagram. On the other hand the diagram including the three-gluon vertex is called the non-Abelian diagram. Comparison of the non-Abelian diagram and the Abelian diagram's contributions makes it possible to dig the differences in the gap equation between the non-Abelian and the Abelian theory.

The method, i.e. going beyond the rainbow approximation by dressing the QGV according to its own equation, has already drawn some attentions~\cite{BDTR,Kra1,FW,WF,AFLS,FNW}.
In Refs. \cite{BDTR}, the authors employed a truncation scheme outlined in Refs.~\cite{Mun,BRS} and took the Munczek-Nemirovsky model~\cite{MN} for the gluon propagator, which is proportional to a Delta function in the momentum space. In such a model, they can iterate the QGV which is a good aspect for using the MN model, but they can only calculate the Abelian contribution directly and they took the non-Abelian contribution into account by rescaling the Abelian part. However, it is the non-Abelian contribution that gives the dominant correction to the QGV~\cite{WF,FNW}, and the rescaling procedure ignored the difference between the kinematics dependence of the non-Abelian part and that of the Abelian part of the QGV. A direct calculation of the non-Abelian contribution is desired and was made in Refs.~\cite{FW,WF} with a Gaussion type interaction model. Those works were more focused on meson observables rather than the DCSB. Besides, the model they used does not respect the ultraviolet (UV) behavior of QCD. It is the major purpose of this work to investigate the impacts of dressing the QGV on the quark propagator and the DCSB (such as consequences on the dynamical quark mass generation, the quark condensate, the critical strength of symmetry breaking and so on.). We are more concentrated on the dominant non-Abelian contribution arising from the three-gluon self interaction, while a comparison to the Abelian contribution is also made. Another important non-Abelian type interaction, the four-gluon interaction, comes into the dressed QGV at two loop level. In this work, we ignored contributions from two and higher loops, so the four-gluon interaction is ignored. To specify, we invoke the equation of motion for the QGV to dress it and make a truncation beyond the rainbow approximation. For the problem to be tractable, the gluon propagator is treated as an input and modeled respecting its UV behavior. We have found that the three-gluon interaction in the QGV makes significant contributions to the dynamical generated quark mass, meanwhile the critical strength of DCSB is overestimated under the rainbow approximation. These results imply that the non-Abelian effects can not be ignored in the dynamical symmetry breaking and employing approximation that drops the non-Abelian feature in QCD may cause large errors.

We also address a technical issue in this work: the renormalization of the gap equation with a dressed QGV. For the method we use, two-loop integrals appear in the gap equation, which generate the overlapping divergences, so it is not a trivial task and deserves a detailed discussion. Refs. \cite{BDTR} used a Delta function type model for the gluon propagator and Refs. \cite{FW,WF} used a Gaussion type model, so both of them are free from the  (UV) divergence, meanwhile both do not respect the UV behavior of QCD. Since the asymptotically freedom is a significant feature of QCD, for completeness, this feature should be realized in a realistic model. In this work we take a model for the gluon propagator respecting its leading log order UV behavior. Quite recently, several papers~\cite{Wil,VW,AW} appeared and improved the studies made in Refs.~\cite{FW,WF} in several aspects including that the UV behavior of the gluon propagator is restored.

This paper is organized as follows. In section 2, we develop a beyond-the-rainbow truncation scheme which deals with the overlapping divergence in the gap equation properly. In section 3, we solve the gap equation numerically and explore the impacts of going beyond the RA. A summary is given in the last section.

\section{The Gap Equation Beyond the Rainbow Approximation}
In this section, we illustrate the beyond-the-rainbow scheme we used for the gap equation. The final result is expressed in Eq. (\ref{DS22}). It picks up the contributions from the three-gluon interaction, and the overlapping divergences are properly removed. The overlapping divergence is well treated in the perturbation theory, however, for the non-perturbative integral equations, it has not been discussed in details. In the perturbation theory, up to a given order, the counter terms (or the renormalization constants) always exactly match the need to cancel the divergences; for the non-perturbative equations, this is not always the case. For example, the equation employed in Ref. \cite{FW} has renormalization constants more than needed if they were using a interaction model that can generate UV divergences. The QGV satisfies several different forms of equations carrying different types and numbers of the renormalization constants, so the equation and the truncation scheme should be chosen carefully so as to make the renormalization constants exactly absorb all the UV divergences. Otherwise, either the divergences are not totally canceled, or the QGV becomes the bare one again (this point will be illustrated in detail later).
\subsection{The Gap Equation}
\begin{figure}[hbt]
\centering
\includegraphics[width = 0.6\textwidth]{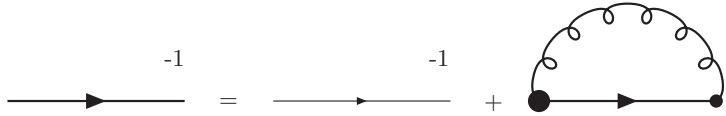}
\caption{The gap equation, i.e. the DS equation for the quark propagator. The bolded lines and vertex are dressed.}\label{gap}
\end{figure}

The gap equation reads~\footnote{We use the Euclidean metric convention in this work. The details of this convention can be found in Ref.~\cite{Rob}, Chapter 4.}
\begin{eqnarray}\label{DS1}
\delta_{\alpha\gamma}S(p,\mu)^{-1}&=&Z_2(\mu,\Lambda)\delta_{\alpha\gamma}\left(i\slashed p+m_b(\Lambda)\right)+Z_{1F}(\mu,\Lambda)\int^\Lambda \frac{d^4 q}{(2\pi)^4}g(\mu)\gamma_\mu T^i_{\alpha\beta}\notag\\
&& \times S(q,\mu)G_{\mu\nu}(p-q,\mu)g(\mu)\Gamma^{i}_{\nu,\beta\gamma}(p,q,\mu),
\end{eqnarray}
where the quark propagator $\delta_{\alpha\beta}S(p,\mu)$, the gluon propagator $\delta_{\alpha\beta}G_{\mu\nu}(p,\mu)$, the quark-gluon vertex $\Gamma^{i}_{\nu,\alpha\beta}(p,q,\mu)$ and the strong coupling $g(\mu)$ are all renormalized quantities; $\mu$ is the renormalization point and $\Lambda$ is an UV cut-off to regularize the theory. $\alpha, \beta, \gamma=1,2,3$ are color indices of the fundamental representation and the superscript $i$ is the color index of the adjoint representation of the color SU(3) group; $T^i$ is the generator. $m_b(\Lambda)$ is the bare quark mass. $Z_2$ and $Z_{1F}$ are the renormalization constants of the quark field and quark-gluon vertex respectively. The renormalization constant for the quark mass $Z_m$ will appear if we express the equation with the renormalized mass $m(\mu)$. It is defined as $Z_2(\mu,\Lambda)m_b(\mu,\Lambda)=Z_4(\mu,\Lambda)m(\mu)=Z_2(\mu,\Lambda)Z_m(\mu,\Lambda)m(\mu)$, where the renormalization constant $Z_4$ has been introduced.
The gap equation can be expressed diagrammatically as shown in Fig. \ref{gap}.

After taking the trace over the color SU(3) space, we have
\begin{eqnarray}\label{DS2}
S(p,\mu)^{-1}&=&Z_2(\mu,\Lambda)\left(i\slashed p+m_b(\Lambda)\right)+\Sigma(p,\mu,\Lambda), \notag\\
\Sigma(p,\mu,\Lambda)&=& Z_{1F}(\mu,\Lambda)C_F\int^\Lambda \frac{d^4 q}{(2\pi)^4} g(\mu)\gamma_\mu S(q,\mu)G_{\mu\nu}(p-q,\mu)g(\mu)\Gamma_{\nu}(p,q,\mu),
\end{eqnarray}
where $\Gamma^{i}_{\nu,\alpha\beta}(p,q,\mu)=T^i_{\alpha\beta}\Gamma_\nu(p,q,\mu)$ has been used. The constant $C_F=\frac{N_c^2-1}{2N_c}$, where $N_c=3$ is the color number. The quark propagator can be written as
$$S(p,\mu)=\frac{Z(p^2,\mu^2)}{i\slashed p+M(p^2)},$$
with $M(p^2)$ called the running quark mass function.

\subsection{The Equation for the Quark-Gluon Vertex}
\begin{figure}[hbt]
\centering
\includegraphics[width = 0.8\textwidth]{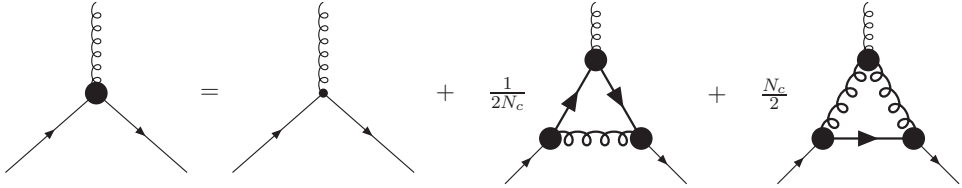}
\caption{The equation of motion for the quark-gluon vertex from the 3-point irreducible effective action. The propagators and vertices in the loops are fully dressed.}\label{qgv1}
\end{figure}

The quark-gluon vertex satisfies its own DS equation, which can be formulated in several different ways~\cite{AFLS}. One can also derive an equation of motion for the QGV from a 3-particle irreducible (3PI) effective action~\cite{Ber}. In principle, one may choose any of them to dress the QGV, and of cause they all need to be truncated in practice. For example,
in the self-consistent three-loop approximation of the 3-point irreducible effective action, the equation for the QGV reads (see Fig. \ref{qgv1})
\begin{equation}\label{DS2.1}
\Gamma_{\nu}(p,q)=Z_{1F}\gamma_\nu+\overline{\Lambda}^{A}_\nu(p,q)+\overline{\Lambda}^{NA}_\nu(p,q),
\end{equation}
where
\begin{equation}\label{DS2.2}
\overline{\Lambda}^A_\nu(p,q)=\frac{g^2}{2N_c}\int\frac{d^4 k}{(2\pi)^4} \Gamma_\beta S(q-p+k)\Gamma_\nu S(k)\Gamma_\alpha G_{\alpha\beta}(p-k),
\end{equation}
\begin{equation}\label{DS4}
\overline{\Lambda}^{NA}_\nu(p,q)=\frac{iN_cg^2}{2}\int\frac{d^4 k}{(2\pi)^4} G_{\alpha\sigma}(p-k)\Gamma_{\nu\sigma\tau}^{3g}(p-k,k-q)G_{\tau\beta}(k-q)\Gamma_{\beta}S(k)\Gamma_{\alpha},
\end{equation}
where we have omitted the renormalization point $\mu$ and the UV cut-off $\Lambda$. $f^{abc}\Gamma_{\nu\sigma\tau}^{3g}(p-k,k-q)$ is the three-gluon vertex with $f^{abc}$ the asymmetric structure constant of SU(3) group. All the internal propagators and vertices are dressed. The superscripts ``$A$" and ``$NA$" denote Abelian diagram (the first triangle loop diagram in Fig. \ref{qgv1}) and the non-Abelian diagram (the second triangle loop diagram in Fig. \ref{qgv1}) respectively.

The Abelian diagram has a color factor proportional to $-1/2N_c$ and the non-Abelian diagram has a color factor proportional to $N_c/2$, so the Abelian contribution is suppressed by a factor of $1/N_c^2$ compared to the non-Abelian one thus is subleading. It is also justified by direct calculations made in Refs.~\cite{WF,AFLS,BT}. In this work, we will check this point later and make further discussions. Now, for simplicity, we ignore the Abelian diagram in the following deduction. A similar expression can be easily obtained for the Abelian diagram. The equation for the QGV now reads:
\begin{equation}\label{DS3}
\Gamma_{\nu}(p,q)=Z_{1F}\gamma_\nu+\overline{\Lambda}^{NA}_\nu(p,q).
\end{equation}
We shall omit the supscript $NA$ on $\overline{\Lambda}_\nu$ to simplify notations.

As mentioned above, for a proper treatment of the renormalization, we need to choose the truncation scheme carefully. To make this point clear, we take a closer look at the truncation scheme used in Ref.~\cite{FW} and compare it with the one originated from Eq. (\ref{DS2.1}). In Ref.~\cite{FW}, the authors take a truncated DSE for the QGV as
\begin{equation}\label{DS-FW1}
\Gamma_{\nu}(p,q)=Z_{1F}\gamma_\nu+Z_{1F}^2Z_1\Lambda_\nu(p,q),
\end{equation}
where $Z_1$ is the renormalization constant of the three-gluon vertex, and $\Lambda_\nu(p,q)$ is Eq. (\ref{DS4}) with all the vertices replaced by the bare ones:
\begin{equation}\label{DS4.1}
\Lambda_\nu(p,q)=\frac{iN_cg^2}{2}\int\frac{d^4 k}{(2\pi)^4} G_{\alpha\sigma}(p-k)\Gamma_{\nu\sigma\tau}^{0,3g}(p-k,k-q)G_{\tau\beta}(k-q)\gamma_{\beta}S(k)\gamma_{\alpha}.
\end{equation}

An important difference of Eq. (\ref{DS-FW1}) to Eq. (\ref{DS3}) is the appearance of the renormalization constants in front of $\Lambda_\nu$. Usually, the divergence of $\Lambda_\nu$ ($\overline{\Lambda}_\nu$) appears as a local divergence in the $\gamma_\nu$ part of the vertex (the meaning of ``local divergence" can be found in the book Ref. \cite{PS}, Chapter 10, Page 337.), and suppose it is $X(\Lambda)\gamma_\nu$, where $X(\Lambda)$ diverges as $\Lambda$ goes to infinity. In the case of Eq. (\ref{DS3}), we have
\begin{equation}\label{DS6}
\Gamma_{\nu}(p,q)=Z_{1F}\gamma_\nu+X(\Lambda)\gamma_\nu+F_\nu(p,q),
\end{equation}
where $F_\nu$ denotes the finite part of $\overline{\Lambda}_\nu$. Renormalization can be easily performed by absorbing $X(\Lambda)$ into $Z_{1F}$.
On the other hand, in the case of Eq. (\ref{DS-FW1}), where $\Lambda_\nu$ comes with a factor $Z_{1F}^2Z_1$, we have
\begin{equation}\label{DS7}
\Gamma_{\nu}(p,q)=Z_{1F}\gamma_\nu+Z_{1F}^2Z_1X(\Lambda)\gamma_\nu+Z_{1F}^2Z_1F_\nu(p,q).
\end{equation}
Since $Z_1$ is the renormalization constant of the three-gluon vertex, it should be determined elsewhere. Suppose it is 1 for simplicity. To fill the requirement that $\Gamma_\nu$ being finite, $Z_{1F}$ has to behave as $X(\Lambda)^{-1/2}$ in the limit $\Lambda\rightarrow \infty$ (up to an irrelevant finite constant) to cancel $X(\Lambda)$, then the first and the third terms in Eq. (\ref{DS7}) vanish and $\Gamma_\nu$ just equals to $\gamma_\nu$ (multiplied by a constant). This is in against with our intention of going beyond the rainbow approximation. So, Eq. (\ref{DS-FW1}) is not adequate when UV divergences appear (which is indeed the case in our study) and we choose Eq. (\ref{DS3}) as our start point for a beyond-the-rainbow scheme. In Ref. \cite{FW}, the model they used makes the integrals convergent, so Eq. (\ref{DS-FW1}) works well in their study.

The three-gluon vertex is taken as an input in this work, and we take it as the bare vertex for simplicity:
\begin{equation}\label{DS5}
\Gamma_{\nu\sigma\tau}^{0,3g}(p-k,k-q)=(q+k-2p)_\tau\delta_{\nu\sigma}+(p+q-2k)_\nu\delta_{\sigma\tau}+(k+p-2q)_\sigma\delta_{\nu\tau}.
\end{equation}

\subsection{The Gap Equation in our Beyond-the-Rainbow Scheme}
Now Eq. (\ref{DS2}), Eq. (\ref{DS3}) and Eq. (\ref{DS4}) form self-contained coupled equations as long as the gluon propagator is given. However, due to the complexity, back-feeding the QGV into its own equation is out of our capabilities now (this difficulty is also recognized in Refs. \cite{AFLS,Wil,VW,AW}). So we need to do a further approximation to make the whole method attainable. This needs be done carefully, because the overlapping divergence in the gap equation should be properly subtracted.

\begin{figure}[hbt]
\centering
\includegraphics[width = 0.2\textwidth]{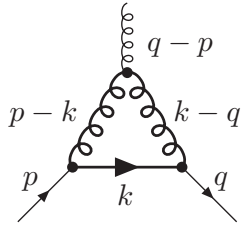}
\caption{The non-Abelian loop diagram $\Lambda_\nu(p,q)$ in Eq. (\ref{DS9}). The propagators are dressed. The vertices are bare ones.}\label{qgv2}
\end{figure}

To this end, we use $Z_{1F}=1+C_{1F}$, where $C_{1F}$ is the counter term, in Eq. (\ref{DS2}) and replace the first $\Gamma_\nu$ with Eq. (\ref{DS3}):
\begin{eqnarray}\label{DS8}
S(p)^{-1}&=&Z_2\left(i\slashed p+m_b\right)+C_F\int \frac{d^4 q}{(2\pi)^4} g^2\gamma_{\mu}S(q)G_{\mu\nu}(p-q)(Z_{1F}\gamma_\nu+\overline{\Lambda}_\nu)\notag\\
&&+C_{1F}C_F\int\frac{d^4 q}{(2\pi)^4} g^2\gamma_\mu S(q)G_{\mu\nu}(p-q)\Gamma_{\nu}(p,q).
\end{eqnarray}
This doesn't change anything. Now we take the approximation by replacing $\Gamma_\nu$ in Eq. (\ref{DS8}) and in $\overline{\Lambda}_\nu$ with the bare one $\gamma_\nu$. Then we have
\begin{eqnarray}\label{DS9}
S(p)^{-1}&=&Z_2\left(i\slashed p+m_b\right)+C_F\int \frac{d^4 q}{(2\pi)^4} g^2\gamma_{\mu}S(q)G_{\mu\nu}(p-q)\gamma_\nu\notag\\
&&+2C_{1F}C_F\int \frac{d^4 q}{(2\pi)^4} g^2\gamma_{\mu}S(q)G_{\mu\nu}(p-q)\gamma_\nu\notag\\
&&+C_F\int \frac{d^4 q}{(2\pi)^4} g^2\gamma_{\mu}S(q)G_{\mu\nu}(p-q)\Lambda_\nu(p,q),
\end{eqnarray}
where $\Lambda_\nu(p,q)$ can be expressed diagrammatically as in Fig. \ref{qgv2}.

Strictly speaking, this approximation breaks the multiplicative renormalizability, which means that the physical quantities and the bare quantities in general no longer $\mu$ independent. However, we checked that the dependence of these quantities on $\mu$ is ignorable in practice. We will be back to this point later.
On the other hand, Eq. (\ref{DS9}) does provide exactly counter terms to cancel the overlapping divergence in the gap equation, and we discuss this point in the next subsection.

In a viewpoint of Feynman diagrams, any truncation scheme can be served as picking up specific (infinite number of) diagrams. The last approximation leading us to Eq. (\ref{DS9}) can be understood in this way and Eq. (\ref{DS9}) specifies our truncation scheme and the diagrams we picked.

\subsection{UV Divergences and the Renormalization}
In this subsection, we discuss the UV divergences in details to show how the divergences (overlapping and overall) be canceled in Eq. (\ref{DS9}). In the integrals, the propagators are all dressed and their UV behavior can be known according to the renormalization group theory. First, for $\Lambda_\nu(p,q)$, by differentiating the integral with respect to one external momentum, it is not difficult to check
that $\Lambda_\nu(p,q)$ has a constant term (which may be divergent) proportional to $\gamma_\nu$ and the remaining terms are finite.
The integral contributing to the constant term behaves as
\begin{equation}\label{DS10}
\sim\int \frac{d^4 q}{q^4}(\ln q)^s= \int \frac{d q}{q}(\ln q)^s,
\end{equation}
at large internal momentum. The integral will diverge if $s\geq -1$.
The remaining part behaves as
\begin{equation}\label{DS11}
\sim\int \frac{d^4 q}{q^5}(\ln q)^{s'}= \int \frac{d q}{q^2}(\ln q)^{s'}= \int \frac{d \ln q}{q}(\ln q)^{s'}=\int d x e^{-x}x^{s'}
\end{equation}
at large internal momentum and always converges for any $s'$.

The UV behavior of the quark propagator up to leading log order is~\cite{GW,Rey}
\begin{equation}\label{DS12}
S(k)\sim\left[\frac{1}{2}\ln(k^2)\right]^{-d_S}\frac{1}{i\slashed k},
\end{equation}
where the quark anomalous dimension
\begin{equation}\label{DS13}
d_S=-4\xi/(33 - 2N_f ).
\end{equation}
$N_f$ is the number of flavors involved in the theory.
To the leading log order, the transverse part of the gluon propagator behaves as
\begin{equation}\label{DS14}
G^{\mathrm{tr}}_{\mu\nu}(q)\sim\left(\delta_{\mu\nu}-\frac{q_\mu q_\nu}{q^2}\right)\frac{1}{q^2}\left[\frac{1}{2}\ln(q^2)\right]^{-d_G},
\end{equation}
where $d_G=(39-9\xi-4N_f )/ [2(33- 2N_f )]$. In this work we use the Landau gauge ($\xi=0$) which is also usually adopted by other DSE studies. In Landau gauge only the transverse part of gluon propagator contributes, so $s=-2d_G-d_S=-(39-4N_f)/(33-2N_f)=-0.92$, where we have taken $N_f=4$.

\begin{figure}[hbt]
\centering
\includegraphics[width = 0.8\textwidth]{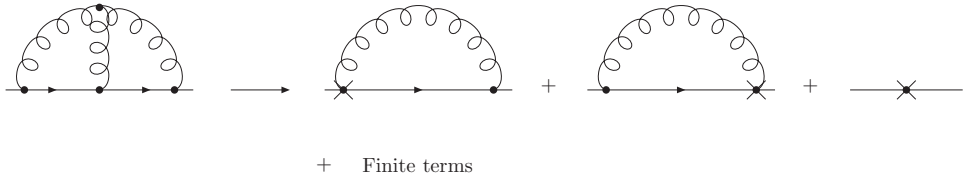}
\caption{Overlapping diagram and its divergent terms. Crosses represent divergent constants.} \label{fig4}
\end{figure}

Finally we find that $\Lambda_\nu(p,q)$ has a similar structure as in the perturbation theory, i.e. it has a term given as $\gamma_\nu$ multiplied by a divergent constant and the remaining part is finite. The divergence can be canceled by the counter term $C_{1F}$. Similar analysis can be made to the integrals in Eq. (\ref{DS9}). The second, the third and the last terms on the r.h.s. of Eq. (\ref{DS9}) all have overall divergences which can be absorbed into $Z_2$ and $Z_4$; in addition, the last term has overlapping divergences as can be seen in Fig. \ref{fig4}. The second term is just what we need to cancel the overlapping divergence and the factor $2$ in front of the $C_{1F}$ plays an important role.

It should be emphasized that since we do not have prior knowledge about the non-perturbative loop integrals, the analysis of the structures of the integrals is necessary. For example, if we choose the number of flavor $N_f$ to be 2 rather than 4, we would have $s=-31/29$, and have no UV divergence in the loop integral.

We need to specify the renormalization conditions for the quark propagator and the QGV to perform the renormalization.
$\Lambda_\nu(p,q)$ can be decomposed into 12 Lorentz invariant functions multiplying independent tensers formed by $\gamma$ matrices and the external momenta:
\begin{equation}\label{DS15}
\Lambda_\nu(p,q)=f_1(p,q)\gamma_\nu+\tilde{\Lambda}_\nu(p,q),
\end{equation}
where $\tilde{\Lambda}_\nu$ denotes all the other terms with other tensers. The renormalization condition for the QGV is chosen to be
\begin{equation}\label{DS16}
Z_{1F}+f_1(p,-p)|_{p^2=\mu^2}=1,
\end{equation}
which implies
\begin{equation}\label{DS17}
C_{1F}=-f_1(p,-p)|_{p^2=\mu^2}.
\end{equation}
The renormalization condition for the quark propagator is
\begin{equation}\label{DS18}
S(p)^{-1}|_{p^2=\mu^2}=i{\slashed p}+m(\mu).
\end{equation}
The quark self energy part $\Sigma(p)$ (see Eq. (\ref{DS2})) can be decomposed as
\begin{equation}\label{DS19}
\Sigma(p)=\Sigma_v(p^2)\slashed p+\Sigma_s(p^2),
\end{equation}
we have
\begin{equation}\label{DS20}
Z_2+\Sigma_v(p^2)|_{p^2=\mu^2}=1,
\end{equation}
and
\begin{equation}\label{DS21}
Z_2m_b+\Sigma_s(p^2)|_{p^2=\mu^2}=m(\mu).
\end{equation}
In the chiral limit, $\Sigma_s$ converges and we do not need Eq. (\ref{DS21}).
After all we have
\begin{eqnarray}\label{DS22}
S(p)^{-1}&=&i\slashed p+m(\mu)-\Sigma_v(\mu^2)i\slashed p-\Sigma_s(\mu^2)+C_F\int \frac{d^4 q}{(2\pi)^4} g^2\gamma_{\mu}S(q)G_{\mu\nu}(p-q)\gamma_\nu\notag\\
&&+C_F\int \frac{d^4 q}{(2\pi)^4} g^2\Big(f_1(p,q)-2f_1(p,-p)|_{p^2=\mu^2}\Big)\gamma_{\mu}S(q)G_{\mu\nu}(p-q)\gamma_\nu\notag\\
&&+C_F\int \frac{d^4 q}{(2\pi)^4} g^2\gamma_{\mu}S(q)G_{\mu\nu}(p-q)\tilde{\Lambda}_\nu(p,q).
\end{eqnarray}
In the chiral limit, we just drop $m(\mu)-\Sigma_s(\mu^2)$ in Eq. (\ref{DS22}). To sum up, the overlapping divergence from the integral including $f_1(p,q)$ is canceled by subtracting $2f_1(p,-p)|_{p^2=\mu^2}$, and there is no other overlapping divergence. In addition all the integrals in Eq. (\ref{DS22}) have overall divergences, and they are canceled by $\Sigma_v(\mu^2)$ and $\Sigma_s(\mu^2)$.

In this section, we presented our beyond-the-rainbow truncation scheme and worked out the renormalization of the gap equation in details. In the previous works~\cite{BDTR,FW,WF}, the authors dressed the QGV with interaction models which make the integrals convergent, so they did not encounter the problem we are facing here.

\section{Numerical Results and Discussions}
\subsection{The Interaction Model}
Eq. (\ref{DS22}) can be solved by a numerical iteration procedure as long as the gluon propagator is treated as an input. In Landau gauge, the gluon propagator can be expressed as
\begin{equation}\label{DS23}
G_{\mu\nu}(k^2)=\left(\delta_{\mu\nu}-\frac{k_\mu k_\nu}{k^2}\right)\frac{D(k^2)}{k^2}.
\end{equation}
The model used in the present work is
\begin{equation}\label{DS24}
g^2\frac{D(k^2)}{k^2}=\frac{4\pi^2Dk^2}{\omega^6}e^{-k^2/\omega^2}+\frac{4\pi^2\gamma_m\ln(\mu^2/\Lambda_{\rm{QCD}}^2)^{d_G-1}}{\left(\frac{1}{2}
\ln\left\{e^2-1+(1+k^2/\Lambda_{\rm{QCD}}^2)^2\right\}\right)^{d_G}}F(k^2),
\end{equation}
\begin{equation}\label{DS25}F(k^2)=\{1-\exp(-k^2/(4m_t^2))\}/k^2,
\end{equation}
where $m_t$, $\omega$ and $D$ are all parameters. We denote this model as Model 1 for convenience, and without indication, calculations are all performed with this Model.
The model's form is inspired by a popular model first used in Ref.~\cite{MT}, which is a DS-BS study under the rainbow-ladder approximation, and is only different from it by the power of the logarithm (if one sets $d_G=1$, the two models would be equivalent). In their case, the model respects the large momentum behavior of the strong coupling $\alpha_s(k^2)$ in accordance of the rainbow approximation, while in our case, the model respects the large momentum behavior of the gluon propagator. We have checked that this difference only causes less than $5\%$ deviations on quantities which we are interested in. So we follow Ref.~\cite{MT} and take $m_t=0.5$ GeV, $N_f=4$, $\Lambda_{\rm{QCD}}^{N_f=4}=0.234$ GeV and $\mu=19$ GeV. In addition it was found that the results from DS-BS studies under the RA are insensitive to $\omega$ in the range $[0.3, 0.5]$ as long as the combination $\omega D$ is a constant~\cite{MT,Qin}, and a typical value of this combination is $\omega D \approx(0.72 ~\rm{GeV})^3$. So we take $\omega=0.5$ GeV and $D=0.74$ GeV$^2$ which give $\omega D \approx(0.72 ~\rm{GeV})^3$. We also need the quark masses as inputs for the explicit chiral symmetry breaking case. We consider the $u$, $d$ and $s$ quarks and follow Ref. \cite{MT} to take
$$m_{u/d}(\mu)|_{\mu=19~\mathrm{GeV}}=3.7~\mathrm{MeV},~~~~~~~~m_s(\mu)|_{\mu=19~\rm{GeV}}=85~\rm{MeV}.$$
For the heavy charm quark, its running mass function is much less effected by the dynamical symmetry breaking, so for our purpose, we do not consider the $c$ quark. $g^2$ also needs specification. If we take $g^2$ at the renormalization point $\mu$, we have
\begin{equation}\label{DS27}
g^2=\frac{4\pi^2\gamma_m}{\ln(\mu^2/\Lambda_{\rm{QCD}}^2)}.
\end{equation}
On the other hand, Eq. (\ref{DS24}) is indeed a model, it is also reasonable to take $g^2=4\pi$ to account for the strong coupling at the low energy region\footnote{Taking $g^2=4\pi$ is suggested by Prof. Richard.Williams (private communication) and we are grateful about this.}. Both of the values are used and compared in this work.

The difference between the model used in Ref.~\cite{MT} and the one used in the present work is related to the ``Abelian approximation" and needs further discussions. In the studies under the rainbow approximation, the ``Abelian approximation" is usually adopted, which replaces the $g^2 D(k^2)$ with the effective running coupling times free gluon propagator $4\pi \alpha(k^2) D^{free}(k^2)$. This replacement is exact in QED, so with this approximation, the gap equation in QCD is the same as that in QED except that the running coupling and the color factor are different. Then the non-Abelian contribution can only get into the equation through the running coupling (and the color factor). The running coupling's behavior can be put in by hand and the large momentum behavior can be restored. Intuitively, it is like QED in an anti-screen medium, while the complexity of the dynamics caused by gauge boson's self interactions is missing. Our study picks up part of the non-Abelian contributions (the three-gluon interaction) through the quark-gluon vertex, so makes an improvement, while non-Abelian contributions from four-gluon interactions, from ghost-gluon interaction which come into the equation through higher order Green functions are still missing. Ref. \cite{MT} took the ``Abelian approximation", so the model they used behaves as $\sim\left[\ln(q^2/\Lambda_{QCD}^2)\right]^{-1}$ at large momentum in accordance with the strong running coupling constant. In this work, we do not take the ``Abelian approximation", so the model we used behaves as $\sim\left[\ln(q^2/\Lambda_{QCD}^2)\right]^{-d_G}$ at large momentum in accordance with the gluon propagator. This is why our model is a little different from the one used in Ref. \cite{MT}.

The first term of the r.h.s. of Eq. (\ref{DS24}) dominates at low momentum and the second term dominates at the large momentum. In the UV limit the model gives
\begin{equation}\label{DS26}
\frac{D(k^2)}{k^2}\rightarrow \frac{1}{k^2}\frac{\ln(\mu^2/\Lambda_{\rm{QCD}}^2)^{d_G}}{\ln(k^2/\Lambda_{\rm{QCD}}^2)^{d_G}},
\end{equation}
which is the asymptotic behavior of the gluon propagator.
If we drop the second term in Eq. (\ref{DS24}), we would have
\begin{equation}\label{DS27.1}
g^2\frac{D(k^2)}{k^2}=\frac{4\pi^2Dk^2}{\omega^6}e^{-k^2/\omega^2}.
\end{equation}
This model is the same as the one used in Ref.~\cite{FW}, and we denote it as Model 2. Model 2 will be considered later for a comparison.

\subsection{Justifying the Renormalization Scheme Explicitly}
We solved Eq. (\ref{DS22}) in the chiral limit and for the $u/d$ quark and the $s$ quark. We shall check explicitly with the numerical results that the renormalization scheme described in the previous section indeed works before proceed to study the physics of the equation. There are two points need to be justified, first, the renormalized quantities should be independent of the UV cut-off and free from UV divergences; and second, physical observables should be independent of the renormalization point $\mu$.

By varying the UV cut-off from $\Lambda^2=10^5$ to $\Lambda^2=10^{11}$, we have observed that, before renormalization, there are indeed UV divergences and after renormalization, the resulted quark propagator is independent of the UV cut-off, thus is free from the UV divergences. The factor 2 in front of $f_1(p,-p)|_{p^2=\mu^2}$ in Eq. (\ref{DS22}) is important to guarantee all the overlapping divergences are removed.
\begin{figure}[hbt]
\centering
\includegraphics[width = 0.6\textwidth]{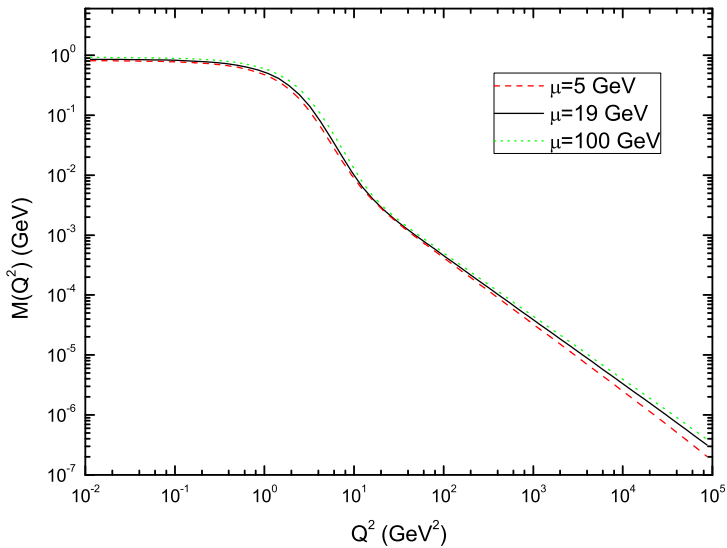}
\caption{The running quark mass at different renormalization points $\mu$.} \label{fig7}
\end{figure}

The independence of the UV cut-off is not sufficient for renormalization, it is also required that the physical observables are independent of the renormalization point $\mu$. At least, for practical purpose, the dependence of observables on the renormalization point $\mu$ should be under control. Strictly speaking, the $\mu$ independence does not hold exactly in the procedure described above. It is the last approximation (which brings us from Eq. (\ref{DS8}) to Eq. (\ref{DS9})) that breaks this feature of renormalization. However, one may expect the results are approximately $\mu$ independent up to some certain precision. We have examined the $\mu$ dependence of some quantities, i.e. pion decay constant, renormalization independent quark condensate, the renormalization independent current quark mass (The expressions of calculating these quantities will be shown later.) as shown in Table \ref{tab2}, and the running quark mass in the chiral limit as shown in Fig. \ref{fig7}, which should be exactly $\mu$ independent in QCD. From the table and the figure, we find that these quantities are insensitive to $\mu$ for large enough $\mu$ ($\mu\gtrsim 19$ GeV). On the other hand, the renormalization point in this work should be chosen large enough to ensure the renormalization constants are approximately flavor independent~\cite{MR}. To sum up, our renormalization scheme does work with an uncertainty (caused by $\mu$ dependence) under control.
\begin{table} \caption{Decay constant $f_\pi$, quark condensate $-\langle\bar{q}q\rangle^0$ and the renormalization independent current quark mass $\hat{m}_{u/d}$ at different renormalization points $\mu$.}
\begin{center}\begin{tabular}{c|c|c|c}
\hline
&$f_{\pi}$(MeV)&$-\langle\bar{q}q\rangle^0$(MeV)$^3$&$\hat{m}_{u/d}$ (MeV) \\
\hline
$\mu=5$ GeV&122&$(250)^3$&$3.9$\\
\hline
$\mu=19$ GeV&123&$(289)^3$&$7.2$\\
\hline
$\mu=100$ GeV&128&$(311)^3$&$7.5$\\
\hline
\end{tabular}\end{center}\label{tab2}
\end{table}

\subsection{The Running Quark Mass $M(Q^2)$ and the Renormalization Function $Z(Q^2)$}
\subsubsection{Impacts of the Three-Gluon Interaction on $M(Q^2)$ and $Z(Q^2)$}
\begin{figure}[htbp]
\begin{minipage}[t]{0.45\textwidth}
\centering
\includegraphics[width = \textwidth]{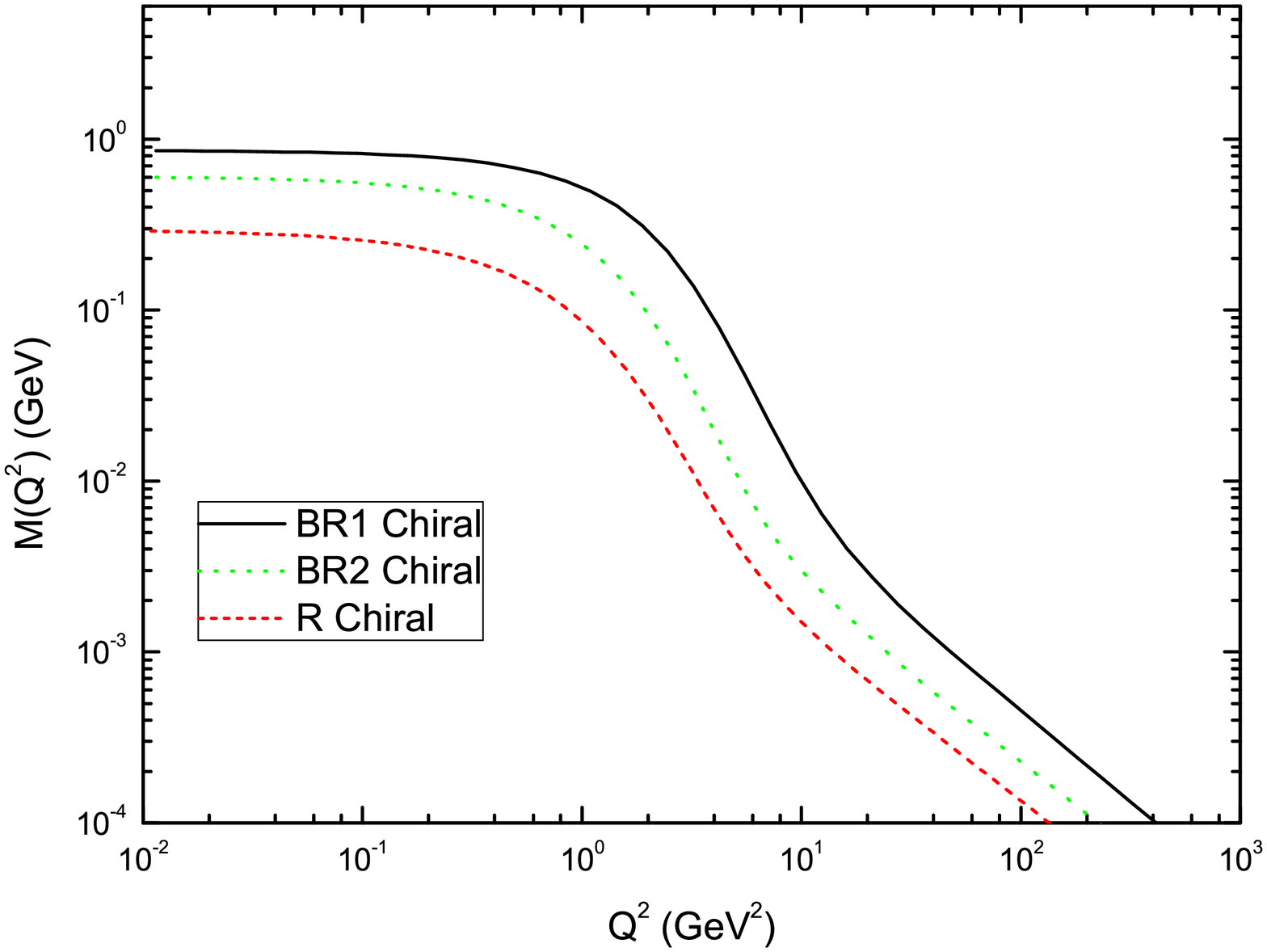}
\caption{The running quark mass function $M(Q^2)$ in the chiral limit. ``BR1" represents beyond-the-rainbow scheme with $g^2$ taken as in Eq. (\ref{DS27}); ``BR2" represents beyond-the-rainbow scheme with $g^2=4\pi$; ``R" represents rainbow approximation.} \label{fig8}
\end{minipage}
\hfill
\begin{minipage}[t]{0.45\textwidth}
\centering
\includegraphics[width = \textwidth]{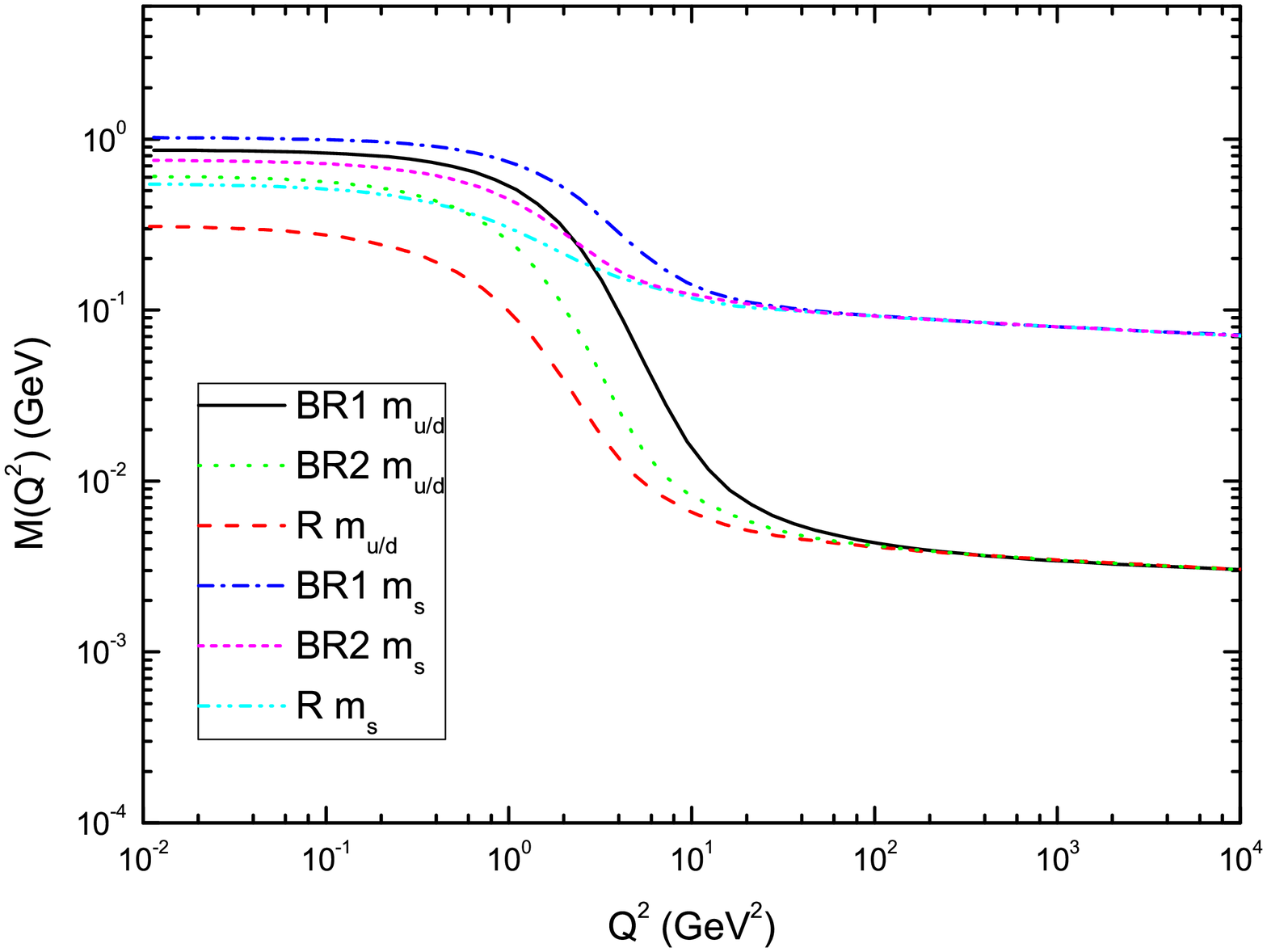}
\caption{The running quark mass function $M(Q^2)$ for $u/d$ quark and $s$ quark. ``BR1" represents beyond-the-rainbow scheme with $g^2$ taken as in Eq. (\ref{DS27}); ``BR2" represents beyond-the-rainbow scheme with $g^2=4\pi$; ``R" represents rainbow approximation.} \label{fig8.1}
\end{minipage}
\end{figure}
\begin{figure}[hbtp]
\begin{minipage}[t]{0.45\textwidth}
\centering
\includegraphics[width = \textwidth]{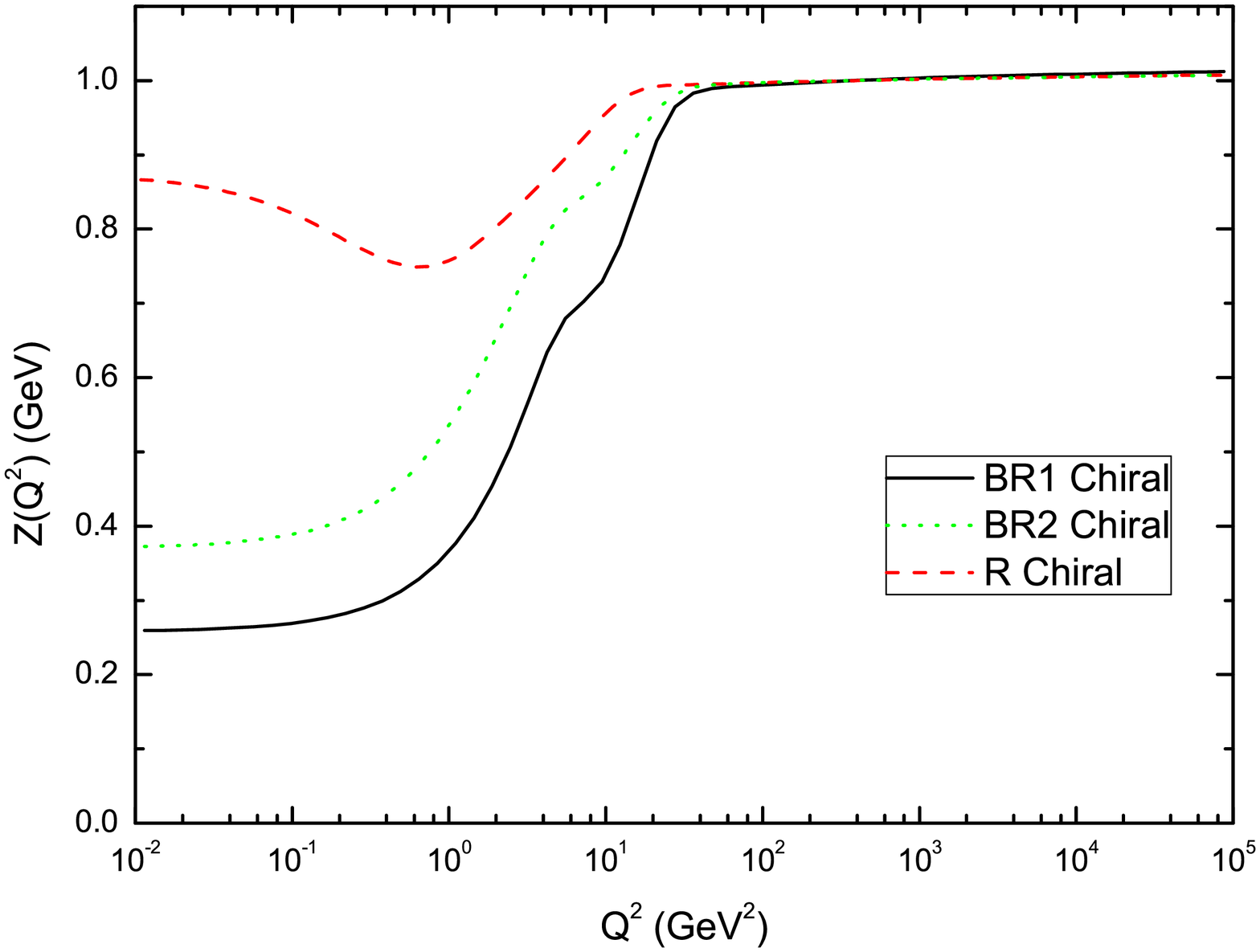}
\caption{$Z(Q^2)$ in the chiral limit. ``BR1" represents beyond-the-rainbow scheme with $g^2$ taken as in Eq. (\ref{DS27}); ``BR2" represents beyond-the-rainbow scheme with $g^2=4\pi$; ``R" represents rainbow approximation.} \label{fig9}
\end{minipage}
\hfill
\begin{minipage}[t]{0.45\textwidth}
\centering
\includegraphics[width = \textwidth]{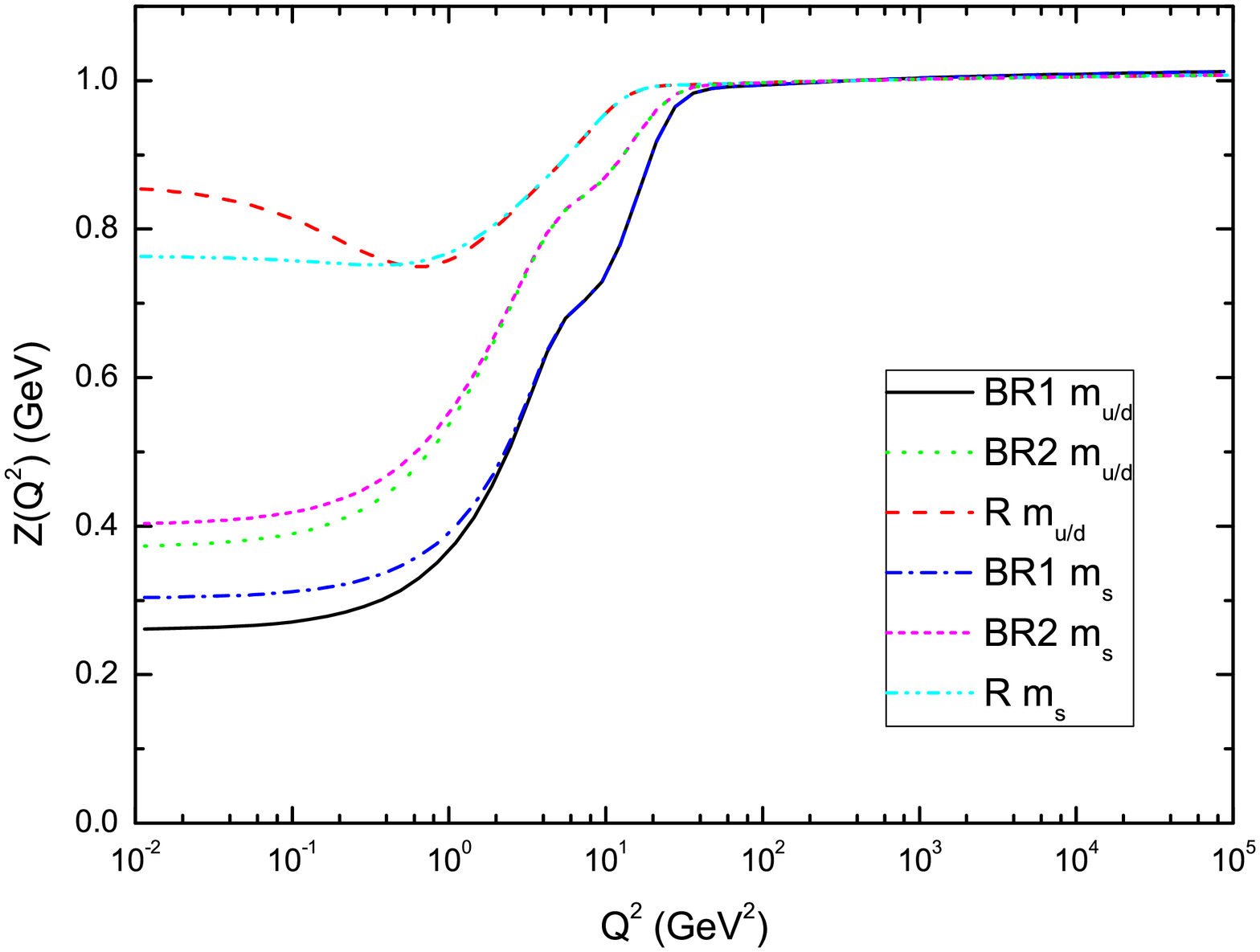}
\caption{$Z(Q^2)$ for for $u/d$ quark and $s$ quark. ``BR1" represents beyond-the-rainbow scheme with $g^2$ taken as in Eq. (\ref{DS27}); ``BR2" represents beyond-the-rainbow scheme with $g^2=4\pi$; ``R" represents rainbow approximation.} \label{fig9.1}
\end{minipage}
\end{figure}
Now we turn to discuss the impacts of dressing the quark-gluon vertex in the gap equation. All the information of the quark propagator is contained in the running quark mass $M(Q^2)$ and the function $Z(Q^2)$. The running quark mass functions in the chiral limit and for $u/d$ quark, $s$ quark are shown in Fig. \ref{fig8} and Fig. \ref{fig8.1}; $Z(Q^2)$ in the chiral limit and for $u/d$ quark, $s$ quark are shown in Fig. \ref{fig9} and Fig. \ref{fig9.1}. ``BR1" and ``BR2" denote the results in the beyond-the-rainbow scheme with $g^2$ taken as in Eq. (\ref{DS27}) and as $g^2=4\pi$ respectively; ``R" denotes the results under RA. It is found that, at the low momentum region, the running quark masses of ``BR's" are as $\sim 2-3$ times large as the results of ``R", and $Z(Q^2)$ of ``BR's" are as $\sim 1/3-1/2$ times large as those of ``R". Our results for the $u/d$ quark are in accordance to those in Ref. \cite{WF}. Dressing the QGV changes the propagator at low momentum region considerably and implies significant impacts on the quantities dominated by the low momentum behavior, such as the pion decay constant, the quark condensate, etc..

Dynamical chiral symmetry breaking takes place in the chiral limit. $M(Q^2=0)$ can be taken as a quantity measuring the dynamical generated quark mass. In the rainbow approximation,
$M(0)=0.29$ GeV; while in the beyond-the-rainbow scheme $M(0)=0.60$ GeV in ``BR2" and $M(0)=0.85$ GeV in ``BR1". Dressing the QGV makes over $50\%$ contributions to dynamical quark mass according to our calculation and these contributions are due to the three-gluon interaction in the QGV part. So using rainbow approximation underestimates the dynamical quark mass.

\subsubsection{Non-Abelian Contributions vs. Abelian Contributions}
\begin{figure}[hbt]
\centering
\includegraphics[width = 0.6\textwidth]{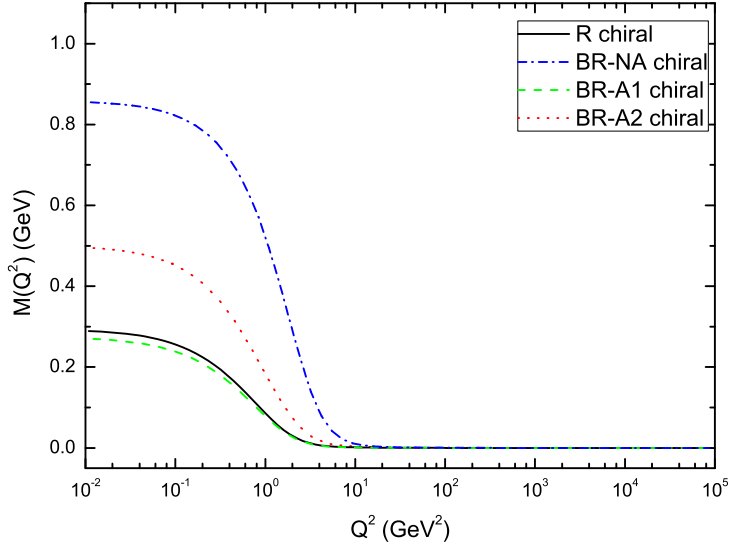}
\caption{Comparisons of non-Abelian contribution and Abelian contribution to $M(Q^2)$. ``BR-NA" represents using $\Gamma_\nu=Z_{1f}\gamma_\nu+\Lambda^{NA}_\nu$ in the beyond the rainbow scheme; ``BR-A1" represents using $\Gamma_\nu=Z_{1f}\gamma_\nu+\Lambda^A_\nu$; ``BR-A2" represents using $\Gamma_\nu=Z_{1f}\gamma_\nu-N_c^2\Lambda^A_\nu$ with the $-N_c^2$ put in by hand.} \label{fig9-2}
\end{figure}

The deviations between the ``BR's" results and ``R" results being so large (at low momentum region) may be understood by observing that the loop diagram to the QGV has two gluon propagators which enhance the interaction strength a lot at low momentum region. And this is due to the non-Abelian three-gluon interaction.

To make this point more clear, we compare the results with non-Abelian diagram contribution to the results with only the Abelian diagram contribution. First, we take $\Gamma_\nu=Z_{1f}\gamma_\nu+\Lambda^A_\nu$ and re-calculate the gap equation beyond the RA. The results are shown in Fig. \ref{fig9-2} denoted as ``BR-A1" (we take the gap equation in the chiral limit as an example). It can be seen from the figure that, the Abelian contribution suppresses the mass function at low energy for that the color factor of the Abelian diagram has opposite sign to the color factor of the non-Abelian diagram. This feature is in accordance to that in Refs. \cite{WF}. Unlike the non-Abelian case, the Abelian contribution modifies $M(Q^2)$ only a little from their RA values and is negligible. This is not a big surprise because the Abelian diagram is $-1/N_c^2$ suppressed compared to the non-Abelian diagram. However this color suppression is not the only reason for this situation, and the difference between the non-Abelian interaction and the Abelian interaction is another reason, as we shall see. To show this, we put a $-N_c^2$ factor in front of the $\Lambda^A_\nu$ by hand, i.e., we take $\Gamma_\nu=Z_{1f}\gamma_\nu-N_c^2\Lambda^A_\nu$, such that the color factors are the same for the non-Abelian and the Abelian diagrams, and re-calculate the gap equation beyond the RA. The resulted $M(Q^2)$ is shown in Fig. \ref{fig9-2} denoted as ``BR-A2". The comparison between ``BR-NA" (with $g^2$ taken as in Eq. (\ref{DS27})) with ``BR-A2" in the figure clearly shows that non-Abelian interaction has great impacts on the gap equation in the beyond-the-rainbow scheme because the interaction is greatly enhanced at the low momentum region due to that the non-Abelian diagram of the QGV includes two gluon propagators in the loop integral.

The non-Abelian diagram is the dominant diagram compared to the Abelian diagram is already known~\cite{WF,AFLS,BT}, however, we make it clear that the difference in the color factors is only one of the reasons. Another important reason is that the non-Abelian diagram includes two gluon propagators which arises from the three-gluon interaction.

\subsubsection{The Large Momentum Behavior}
\begin{figure}[hbt]
\centering
\includegraphics[width = 0.6\textwidth]{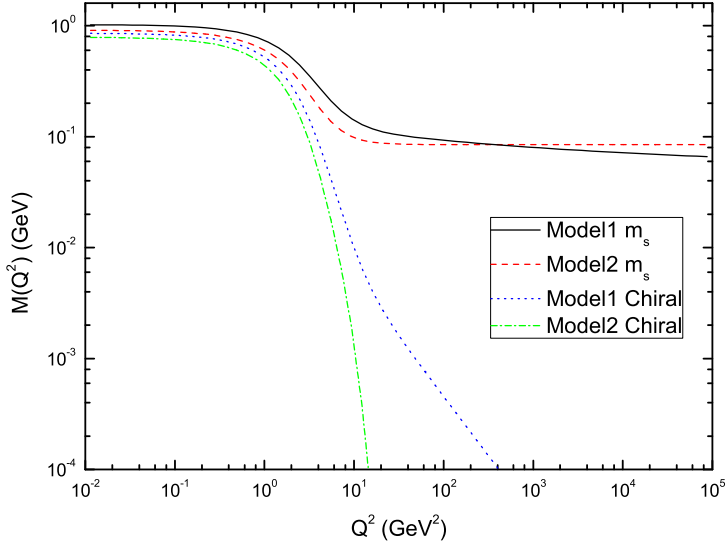}
\caption{Comparing $M(Q^2)$ resulted from Model 1 and Model 2.} \label{fig9-1}
\end{figure}

At the large momentum region, since we use an interaction model with correct UV behavior of the gluon propagator, our results should reproduce the correct UV behavior for the quark propagator (approximately, not exactly, because the beyond-the-rainbow gap equation is not the exact equation for the quark propagator).
The running quark mass at large momentum in the chiral limit behaves as~\cite{MR}
\begin{equation}\label{DS30}
M(p^2) \xrightarrow[]{\mathrm{large~} p^2}\frac{2\pi^2\gamma_m(-\langle\bar{q}q\rangle^0)}{3p^2\left[\frac{1}{2}\ln(p^2/\Lambda_{QCD}^2)\right]^{1-\gamma_m}},
\end{equation}
where $\langle\bar{q}q\rangle^0$ is the renormalization point independent quark condensate. In the explicit chiral symmetry breaking case, it is dominated by the current quark mass and behaves, at one loop order of the renormalization group equations, as
\begin{equation}\label{DS31}
M(p^2) \xrightarrow[]{\mathrm{large~} p^2}\frac{\hat{m}}{\left[\frac{1}{2}\ln(p^2/\Lambda_{QCD}^2)\right]^{\gamma_m}},
\end{equation}
where $\hat{m}$ is the renormalization independent current quark mass. $\gamma_m=12/(33-2N_f)$ is the anomalous dimension at this order. We can see from Fig. \ref{fig8} and Fig. \ref{fig8.1} that our results are consistent with these two UV behavior at large momentum region. It would be interesting to compare the results with the interaction model having the correct UV behavior to the one not having it. So we also calculated the gap equation with Model 2 described by Eq. (\ref{DS27.1}). In Fig \ref{fig9-1}, we show $M(Q^2)$ resulted from Model 1 and Model 2 for comparison. For Model 2, $M(Q^2)$ in the chiral limit falls off to 0 obviously faster than $1/p^2$ as $p^2\rightarrow\infty$; and $M(Q^2)$ for $m_s$ quark becomes flat at large momentum region. They clearly do not obey the behavior described by Eqs. (\ref{DS30}) and (\ref{DS31}), thus one can not extract the renormalization point independent quark condensate and current quark mass from them.

We have another interesting observation. At large momentum region, one may expect that the perturbation theory applies due to the asymptotic freedom so that results from ``BR" and ``R" approach each other. However in the chiral limit case, the quantity $Q^2 M(Q^2)$ from the beyond-the-rainbow scheme and the rainbow approximation remain different at large momentum (see Fig. \ref{fig8}), which indicates that non-perturbative effects may also be revealed at large momentum region. This feature has not been mentioned in previous studies.

\subsection{Quark Condensate, Pion Decay Constant and Pion Mass}
It is important to discuss the effects of dressing the QGV on the quark condensate, which measures the dynamical symmetry breaking, and some typical physical observables. We consider the pion decay constant and the pion mass here. The pion decay constant can be calculated approximately using the equation given by~\cite{RW},
\begin{equation}\label{DS28}
f_\pi^2=\frac{3}{4\pi}\int dp^2\frac{p^2Z(p^2)M(p^2)}{(p^2-M^2(p^2))^2}\left(M(p^2)-\frac{p^2}{2}\frac{dM}{dP^2}\right),
\end{equation}
which followed from making an assumption on the pseudoscalar vertex's structure in the soft pion limit and the chiral limit. $f_\pi$ calculated with this equation under the rainbow approximation is $75$ MeV, while the $f_\pi$ obtained with the (almost) same parameters in Ref. \cite{MT} gave the value in accordance with the experimental value $f_\pi^{\mathrm{exp.}}\simeq 92$ MeV. So Eq. (\ref{DS28}) underestimates $f_\pi$ for about $\sim17$ MeV (It was also indicated in Ref. \cite{FA} that Eq. (\ref{DS28}) underestimates $f_\pi$). The error due to using Eq. (\ref{DS28}) does not provide any interesting information. Since we are interested in comparing the beyond-the-rainbow results with the RA results, we add up with $17$ MeV on all the $f_\pi$ calculated in the BR scheme and in the RA scheme, so that the $f_\pi$ under RA is re-adjusted to be the experimental value.

The quark condensate at the renormalization point $\mu$ can be obtained by integrating the propagator's scalar part over the momentum and we have
\begin{equation}\label{DS29}
-\langle\bar{q}q\rangle^0_\mu \equiv N_c Z_4 \int\frac{d^4p}{(2\pi)^4}\mathrm{Tr}\left[S(p,\mu)\right].
\end{equation}
The supscript ``0" indicates the quantity is taken in the chiral limit. The renormalization point independent quark condensate is extracted from the data of running quark mass with Eq. (\ref{DS30}).
Having the decay constant and the quark condensate, one can obtain the mass of pion from the Gell-Mann-Oakes-Renner relation up to $\mathcal{O}(\hat{m})$ order.
\begin{table} \caption{Comparing the pion mass, pion decay constant, the quark condensate at the renormalization point and the renormalization point independent quark condensate between the beyond-the-rainbow scheme (denoted as ``BR1" and ``BR2") and the rainbow approximation scheme (denoted as ``R").}
\begin{center}\begin{tabular}{c|c|c|c|c}
\hline
&$f_{\pi}$(MeV)&$m_{\pi}$(MeV)&$-\langle\bar{q}q\rangle^0_{\mu=19\rm{GeV}}$(MeV)$^3$ &$-\langle\bar{q}q\rangle^0$(MeV)$^3$\\
\hline
BR1&123&152&$(369)^3$&$(289)^3$\\
\hline
BR2&105&157&$(301)^3$&$(266)^3$\\
\hline
R&92&143&$(254)^3$&$(229)^3$\\
\hline
\end{tabular}\end{center}\label{tab3}
\end{table}

Comparisons of $f_\pi$, $m_\pi$ and the quark condensates between the RA scheme and the BR scheme are shown in Table \ref{tab3}. It is found that dressing the QGV enhances these quantities a lot as compared to the results under RA scheme, which justifies the conclusions about the quark propagators on the same topic. Quantitatively, the decay constant is enhanced by $\sim 13-31$ MeV as $g^2$ varies. This result is in accordance with Ref.~\cite{FW}. Similar conclusions may be drawn to the quark condensate.

\begin{figure}[hbt]
\centering
\includegraphics[width = 0.6\textwidth]{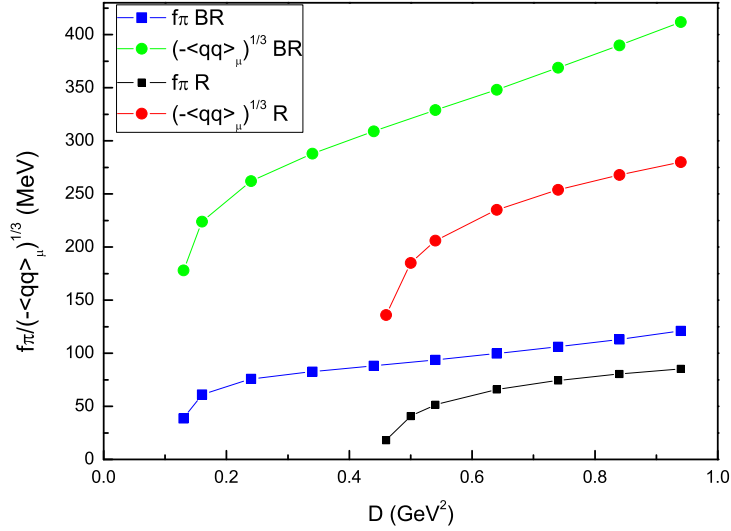}
\caption{Quark condensate and the pion decay constant vs $D$, i.e. the strength of the interaction at low momentum.} \label{fig10}
\end{figure}

The large enhancement of the quark condensate implies a large suppression of the critical value of the interaction strength for dynamical symmetry breaking. In the model used here, the parameter who plays the role as the strength of the interaction in the low momentum region is the parameter ``$D$", which has been taken to be $D=0.74$ GeV$^2$. Now we vary this parameter (with $g^2$ taken as in Eq. (\ref{DS27})) and exhibit the quark condensate and the pion decay constant, both of which characterize the dynamical chiral symmetry breaking in the chiral limit, at different $D$'s in Fig. \ref{fig10}. It can be seen from the figure that the critical value of having the dynamical symmetry breaking moves from $D\simeq0.45$ GeV$^2$ in the rainbow approximation to $D\simeq 0.12$ GeV$^2$ in the beyond-the-rainbow scheme. And in the region where $D$ is between the two critical points, dressing the QGV has essential effects.

\begin{figure}[hbt]
\centering
\includegraphics[width = 0.6\textwidth]{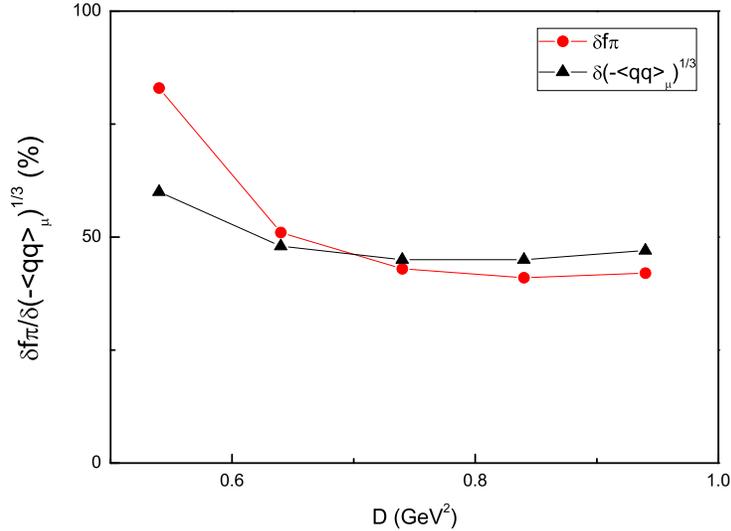}
\caption{Relative deviations (i.e. $\delta X\equiv\frac{X_{BR}-X_{R}}{X_{R}}$ where $X_{R(BR)}$ represents quantity under the RA(BR) scheme) of pion decay constant and the quark condensate vs $D$.} \label{fig11}
\end{figure}

Another observation is that, for $D$ sufficiently larger than $0.45$ GeV$^2$, the relative deviations of $f_\pi$ and the quark condensate (actually the cuberoot of minus quark condensate) in the BR scheme from those in the RA scheme are roughly constant with respect to $D$, which can be seen explicitly from Fig. \ref{fig11}. This feature means that the contributions of dressing the QGV (due to the three-gluon interaction) in the gap equation will not approach to zero as the strength of interaction varies as long as the chiral symmetry is dynamically broken. What we would like to stress is that, dressing the QGV has essential impacts on the dynamical chiral symmetry breaking and the quantities relevant to it.

\section{Summary}

We studied the quark propagator and some relevant quantities with the gap equation under a beyond-the-rainbow scheme, and an interaction model respecting the asymptotic freedom behavior of QCD is used. The gap equation has divergent integrals and the renormalization is performed. Renormalizing the gap equation in the beyond-the-rainbow truncation scheme is not a trivial task due to the equation's non-perturbative nature and the appearance of overlapping divergences. The important point to solve this problem is that the truncation scheme needs to be chosen carefully to allow a suitable subtractive renormalization.

With this method, we analyzed the impacts of going beyond the rainbow approximation. It is found that three-gluon interaction contributions to the QGV in the gap equation changes the quark propagators significantly. The three-gluon interaction makes over $50\%$ contributions to the dynamical quark mass; the pion decay constant is enhanced by $\sim 13-31$ MeV. We have taken different parameters ($g^2$ and $D$) to show how the results related to model parameters. In a large region of parameters, the three-gluon interaction in the QGV makes important contributions. It is also found that going beyond the rainbow approximation lowers the critical strength of dynamical chiral symmetry breaking, which means using rainbow approximation overestimates the critical strength of dynamical chiral symmetry breaking. These results imply that the non-Abelian effects can not be ignored in the dynamical symmetry breaking and employing approximation that drops the non-Abelian feature in QCD may cause large errors. The comparison of the non-Abelian contribution and the Abelian contribution indicates that non-Abelian contribution is dominant even after cutting off the effect of the color factor, the reason is that the three-gluon interaction contributes two gluon propagators in the loop integral of the QGV, and largely enhances the low energy strength.

We calculated hadron observables with approximated expressions without invoking the BS equation, which limited our abilities of giving faithful values of these quantities and exploring more observables. In addition, a complete non-perturbative treatment of the QGV requires feeding-back the vertices in the equations, which is not performed in this study due to technical difficulties. Further efforts should be devoted in these directions.

\section*{Acknowledgments}
This work was supported by the National Science Foundation of China (NSFC) under Grant No. 11475092,  the Specialized Research Fund for the Doctoral Program of High Education of China No.20110002110010, and the Tsinghua University Initiative Scientific Research Program No.20121088494.


\begin{thebibliography}{0}
\bibitem{AJ} D. Atkinson and P. W. Johnson, Phys. Rev. D 37, (1988) 2290; Phys. Rev. D 37, (1988) 2296.
\bibitem{MJ} H. Z. Munczek and P. Jain, Phys. Rev. D 46, (1992) 438; Phys. Rev. D 48, (1993) 5403.
\bibitem{RW} C. D. Roberts and A. G. Williams, Prog. Part. Nucl. Phys. 33, (1994) 477.
\bibitem{MR} P. Maris and C. D. Roberts, Phys. Rev. C 56, (1997) 3369.
\bibitem{MT} P. Maris and P. C. Tandy, Phys. Rev. C 60, (1999) 055214.
\bibitem{CM} S. R. Cotanch and P. Maris, Phys. Rev. D 66, (2002) 116010.
\bibitem{MRRS} P. Maris, A. Raya, C. D. Roberts and S. M. Schmidt, Eur. Phys. J. A 18, (2003) 231.
\bibitem{Kra} A. Krassnigg, Phys. Rev. D 80, (2009) 114010; V. Mader, G. Eichmann, M. Blank and A. Krassnigg, Phys. Rev. D 84, (2011) 034012; T. Hilger, C. Popovici, M. G$\acute{\mathrm{o}}$ez-Rocha and A. Krassnigg, Phys. Rev. D 91, (2015) 034013.
\bibitem{BDTR} A. Bender, W. Detmold, A. W. Thomas and C. D. Roberts, Phys. Rev. C 65, (2002) 065203; M. S. Bhagwat, A. H$\ddot{\mathrm{o}}$ll, A. Krassnigg, C. D. Roberts and P. C. Tandy, Phys. Rev. C 70, (2004) 035205; H. H. Matevosyan, A. W. Thomas and P. C. Tandy, Phys. Rev. C 75, (2007) 045201.
\bibitem{Kra1} M. G$\acute{\mathrm{o}}$ez-Rocha, T. Hilger and A. Krassnigg, Phys. Rev. D 92, (2015) 054030.

\bibitem{FW}C. S. Fischer and R. Williams, Phys. Rev. Lett. 103, (2009) 122001.
\bibitem{WF}R. Williams and C. S. Fischer, Chin. Phys. C 34, (2010) 1500; R. Williams, EPJ Web of Conferences 3, (2010) 03005.
\bibitem{RHK} M. G$\acute{\mathrm{o}}$mez-Rocha, T. Hilger and A. Krassnigg, Few-Body Syst. 56, (2015) 475.

\bibitem{BC} J. S. Ball and T.-W. Chiu, Phys. Rev. D 22, (1980) 2542.
\bibitem{CP1} D. C. Curtis and M. R. Pennington, Phys. Rev. D 42, (1990) 4165.
\bibitem{CP2} D. C. Curtis and M. R. Pennington, Phys. Rev. D 44, (1991) 536; Phys. Rev. D 46, (1992) 2663; Phys. Rev. D 48, (1993) 4933.
\bibitem{HW2} F. T. Hawes and A. G. Williams, Phys. Lett. B 268 (1991) 271.
\bibitem{HW} F. T. Hawes and A. G. Williams, Phys. Rev. D 51, (1995) 3081.
\bibitem{HMR} F. T. Hawes, P. Maris and C. D. Roberts, Phys. Lett. B 440, (1998) 353.
\bibitem{Cha} L. Chang, {\it et al.}, Phys. Rev. C 79, (2009) 035209.
\bibitem{CR} L. Chang and C. D. Roberts, Phys. Rev. Lett. 103, (2009) 081601.
\bibitem{FA} C. S. Fischer and R. Alkofer, Phys. Rev. D 67, (2003) 094020.
\bibitem{AP} A. C. Aguilar and J. Papavassiliou, Phys. Rev. D 83, (2011) 014013.
\bibitem{Roj} E. Rojas, {\it et al.}, J. High Energ. Phys. 1310, (2013) 193.
\bibitem{He} H.-X. He, Phys. Rev. D 80, (2009) 016004.
\bibitem{QCLRS} S.-X. Qin, {\it et al.}, Phys. Lett. B 722 (2013) 384.

\bibitem{AFLS}R. Alkofer, C. S. Fischer, F. J. Llanes-Estrada and K. Schwenzer, Annals of Physics 324, (2009) 106.
\bibitem{FNW}C. S. Fischer, D. Nickel and J. Wambach, Phys. Rev. D 76, (2007) 094009; C. S. Fischer and R. Williams, Phys. Rev. D 78, (2008) 074006; C. S. Fischer, D. Nickel and R. Williams, Eur. Phys. J. C 60, (2009) 47.



\bibitem{Mun}H. J. Munczek, Phys. Rev. D 52, (1995) 4736.
\bibitem{BRS} A. Bender, C. D. Roberts, and L. Von Smekal, Phys. Lett. B 380, (1996) 7.
\bibitem{MN} H.J. Munczek and A.M. Nemirovsky, Phys. Rev. D 28 (1983) 181.

\bibitem{Wil} R. Williams, Eur. Phys. J. A 51, (2015) 57.
\bibitem{VW} M. Vujinovic and R. Williams, Eur. Phys. J. C 75, (2015) 100.
\bibitem{AW} H. Sanchis-Alepuz and R. Williams, Phys. Lett. B 749, (2015) 592.



\bibitem{Rob} C. D. Roberts, {\it Primer for Contemporary Quantum Field
Theory in Hadron Physics at Nonzero Temperature and Density}, Lecture Notes for a Graduate Level Course
given in {\it Fachbereich Physik, Universit$\ddot{a}$t Rostock}:
Winter Semester, 2001; Spring Semester, 2002.
\bibitem{Ber}J. Berges, Phys. Rev. D 70, (2004) 105010.

\bibitem{BT}M. S. Bhagwat and P. C. Tandy, Phys. Rev. D 70, (2004) 094039.
\bibitem{PS}M. E. Peskin and D. V. Schroeder, An Introduction to Quantum Field Theory, Publisher: Westview (1995).
\bibitem{GW} D. J. Gross and F. Wilczek, Phys. Rev. D 8, (1973) 3633.
\bibitem{Rey} E. Reya, Phys. Rept. (Review Section of Physics Letters) 69, (1981), 195.
\bibitem{Qin} S.-X. Qin, Phys. Rev. C 84, (2011) 042202(R).


\end{thebibliography}
\end{document}